*Chronicle*

# "Foot of the iceberg" of Nobel Prize in Physics 2025: ILTPE and LTP contribution

O. G. Turutanov[a]

B. I. Verkin Institute for Low Temperature Physics and Engineering of the National Academy of Sciences of Ukraine, 47 Nauky Ave., 61103 Kharkiv, Ukraine

[a]Present address: Comenius University, Mlynska dolina F2, 84248 Bratislava, Slovakia.

This year, the Nobel Prize 2025 was awarded to an international team of two experimentalists, British-born John Clarke and American John M. Martinis, and a theoretician from France, Michelle H. Devoret, all worked together at the University of California, "for the discovery of macroscopic quantum tunneling and the quantization of energy in an electrical circuit" [1].

Quantum mechanics has recently turned 100, but the potential of its revolutionary ideas is far from exhausted. In 1957, the BCS quantum mechanical theory of superconductivity was developed, for which the authors, John Bardeen, Leon Cooper, and John Robert Schrieffer, also received the Nobel Prize in 1972. According to it, a single Bose condensate with a common complex wave function is formed in a superconductor due to the weak attraction between electrons caused by the phonon exchange. The coherence of this wave function is preserved even if the superconductor is interrupted by a thin dielectric layer, which was in doubt from the point of view of many researchers at the time until Brian Josephson theoretically proved this in 1962. Unexpectedly, the probability of the Cooper pair tunneling turned out to be anomalously high, almost the same as for single electrons. (Cooper pairs, as one might simplistically assume, constitute a superconducting condensate; although, strictly speaking, the BCS theory considers many-particle interactions.) This phenomenon is known as the stationary Josephson effect. According to his theory, the maximum possible superconducting current through a tunnel junction is proportional to the sine of the condensate wave function phase difference between the two parts of the superconductor. Thus, the phase of the wave function transformed from a purely theoretical quantity into a real one measurable through the superconducting current. Josephson also derived a second, non-stationary, effect that relates the frequency of current oscillations in such a junction to the constant voltage across it, if the critical current of the junction is exceeded. For these two predictions, he received the Nobel Prize in 1973. Shortly after, both effects were confirmed experimentally. Note that extremely weak picowatt radiation from a Josephson junction with a frequency of about 10 GHz (three-centimeter frequency range, X-band) was first experimentally observed at the B. I. Verkin Institute for Low Temperature Physics and Engineering of the National Academy of Sciences of Ukraine (ILTPE) by Ukrainian scientists I. K. Yanson, V. M. Svistunov, and I. M. Dmitrenko in 1965 [2]. Almost simultaneously, two months later, the American team D. N. Langenberg, D. J. Scalapino, B. N. Taylor, and R. E. Eck published its results [3].

A group of American researchers and engineers from Ford Research Labs developed two new superconducting devices: a DC SQUID with two Josephson junctions (R. C. Jaklevic, J. Lambe, A. H. Silver, and J. E. Mercereau, 1964) and a single-junction RF SQUID (J. E. Zimmerman and A. H. Silver, 1967). In the latter, the junction is closed by a superconducting loop (ring), and the magnetic flux inside the loop is proportional to the wave function phase difference across the Josephson junction. This further simplified the measurement of the phase as a real physical quantity, and SQUIDs became perhaps the most "popular" superconducting devices from that time to the present, and the base for all modern superconducting electronics. It includes quantum applications, such as superconducting qubits (quantum bits) for quantum processors and single-photon sensors for secure communication systems.

The Josephson junction and the SQUID are macroscopic systems, both in size and in the number of particles contained in them. Moreover, they are characterized by macroscopic quantum states, which involve all the particles that create the superconducting condensate. But these systems fundamentally differ from microparticles in that they have a strong electromagnetic coupling with the dissipative environment. Therefore, it would seem that one should not expect such quantum phenomena as tunneling of macroscopic quantum states in them, because they would be damped by dissipation, which destroys the wave function coherence. Moreover, dissipation has been largely absent in the quantum mechanics of microparticles so far. Nevertheless, in 1978, Anthony Leggett (also a 2003 Nobel laureate) proposed a theoretical idea about the possibility of tunneling of macroscopic quantum states (MQT) [4] and later, together with his graduate student A. O. Caldeira, developed a detailed theory of this process in dissipative systems [5, 6]. Other theorists also expressed similar ideas, but MQT is "officially" associated with Leggett's name. As a whole, this created a reliable basis for setting up relevant experiments.

Here, one can find some analogy with Josephson's discovery and the logic of scientific thought in general. Josephson showed that the probability of two-particle tunneling (Cooper pair tunneling) is almost of the same order of magnitude as that of one-particle tunneling, due to the correlation between two electrons in a Cooper pair. Leggett advanced further along this way, stating a high probability of tunneling of a macroscopic quantum variable, the phase or magnetic flux in SQUID, which defines a common coherent state of a large number of correlated particles, provided that the dissipation in the system is not very large. The latter means the necessity of isolating the system from the electromagnetic environment in the entire frequency band and the usage of ultralow millikelvin temperatures, at which the energy of thermal fluctuations is much lower than the distance between the quantum levels of the system. These requirements create great problems for experimenters.

It is also necessary to emphasize the difference between the discoveries of Josephson and the current Nobel laureates, together with Leggett. In the first case, Josephson considered the tunneling of *individual* Cooper pairs, which are *microparticles*, rather than the tunneling of the entire superconducting condensate, although the superconducting banks were in macroscopic quantum states. Leggett considered the *simultaneous* tunneling of the *macroscopic* quantum state *as a whole* as a "macroscopic degree of freedom", or a macroscopic variable, and not as a sequence of tunneling microparticles, which form this common state. And Clark, Martinis, and Devore proved this experimentally. Therefore, these are really different physical achievements for which the corresponding Nobel Prizes were awarded.

A well-designed and well-executed experiment was performed by the present Nobel laureates [7] on a single Josephson junction cooled down to a temperature of about 20 mK. Short pulses of the current slightly below the critical value were passed through it, and the frequency of spontaneous transitions to the resistive state (indicated by voltage pulses) was measured. The measurements were successful thanks to the experimental skill of Martinis, who was then a graduate student. The experiment showed that the macroscopic degree of freedom, which was here the phase difference across the Josephson junction, associated with the entire superconducting condensate, obeys the quantum mechanical laws. This result was one of those that led to the Nobel Prize. Further experiments, in particular the microwave spectroscopy (the absorption of microwave radiation), proved the presence of discrete quantized energy levels in the macroscopic system [8].

Meanwhile, Ukrainian scientists Georgiy Tsoi and Vladimir Shnyrkov, working at the ILTPE under the guidance of Igor Dmitrenko, observed MQT already in 1981 in an RF SQUID [9] keeping in mind the same Leggett's ideas [4–6]. Nobel laureates, in their mentioned-above work [7], cited only the more recent paper [10] of the ILTPE researchers. However, this work was really better due to the essential theoretical part by the talented theorist Viktor Khlus. In this series of papers, the magnetic flux piercing the SQUID loop (almost the same as the phase difference across the Josephson contact) reflected the collective behavior of a great set of microparticles with a macroscopic net mass, but, anyway, they obeyed the quantum mechanical laws. Totally measurable magnetic flux could change as a result of quantum tunneling! In addition, the SQUID experiments had an advantage over the work of Martinis *et al.* [7] with their Josephson contact, which would pass to a finite-voltage state, thus dissipating heat. The SQUID did not go into a resistive state, and therefore, there was no need to wait a long time for thermalization (cooling) of the contact after detecting the voltage pulse. One should also keep in mind that these experiments were long-running and required special precautions to get rid of EMI (electromagnetic interference). It should be noted that the whole picture of experiments in this field, the history of interim results, and the struggle of ideas in reality was highly complex and involved the contributions of many scientists. Let us mention, for example, the names of Rudolf de Bruyn Ouboter and Terrence D. Clark among them.

Figure 1 illustrates the experimental results of the MQT observation. The figure displays high-frequency current-voltage characteristics (RF IVCs) of a SQUID along with their derivatives. The derivatives feature small additional maxima at the values of the magnetic flux inside the loop, which make the potential symmetric. In this situation, at a low enough temperature, one can observe non-dissipative MQT transitions between two potential wells (two magnetic or current states) of the superconducting loop.

Thus, one of the cornerstones of the science we are discussing has been established: experimental proof of the existence of macroscopic quantum tunneling and energy quantization in macroscopic systems.

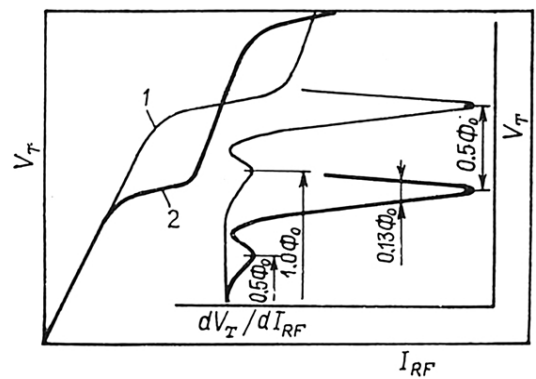

*Fig. 1.* RF IVC of a SQUID and their derivatives at $T \simeq 0.5$ K, $l \simeq 7.5$ ($\varphi_{ec} \simeq 3\pi$) and the values of the external magnetic flux $\varphi_e = 2\pi n$ (*1*) and $\varphi_e = 2\pi(n + \frac{1}{2})$ (*2*). [Fig. 7 from *Sov. J. Low Temp. Phys.* **11**, 77 (1985)].

The role of the *Fizyka Nyzkykh Temperatur/Low Temperature Physics* (*FNT/LTP*) journal should be outlined here. It was founded in the ILTPE in 1975 and celebrated its successful 50th anniversary this year, keeping its strong position as one of the best Ukrainian scientific journals. Since its founding, the journal has become one of the central monthly editions in the field of low-temperature physics, publishing works of a very high level. Thanks to the publication of the English version of the *Fizyka Nyzkykh Temperatur* by the American Institute of Physics (AIP) as *Low Temperature Physics*, scientists worldwide can learn about the research conducted at the ILTPE and worldwide. This long-standing publication tradition also includes the above-mentioned works [10] that later resonated with future Nobel laureates, highlighting the journal's significant contribution to the development of modern low-temperature physics.

By reducing the dissipation in a macroscopic quantum system by two or three orders of magnitude, one can come to the observation of a more subtle and extremely important phenomenon, the superposition of macroscopic quantum states, which is an even more amazing quantum property in relation to macroscopic systems. The simultaneous coexistence of two or more quantum states in a system opens up the possibility of creating a fundamentally new information storage element, a qubit, and further operations with quantum information. The field is now called quantum engineering. Since superposition requires much less dissipation, many considered it impossible, unlike MQT. Ukrainian researchers from ILTPE managed to observe this phenomenon earlier than others [11, 12], but at that time, most theorists did not understand these experimental results, and, as often happens in science, the world priority was assigned to other scientists [13, 14].

In conclusion, one can say that the Nobel Committee, in awarding the prize this year, 2025, honored not only outstanding achievements in basic physics but also their enormous importance for practical implementation, which has already made the 21st century mostly quantum. It is a pleasure to realize that Ukrainian scientists were among the first to make their contribution to this global process.

The author is grateful to V. I. Shnyrkov, Yu. G. Naidyuk, O. V. Dolbin, and K. M. Matsievskii for the provided information and fruitful discussions.

1. https://www.nobelprize.org/prizes/physics/2025/summary/
2. I. K. Yanson, V. M. Svistunov, and I. M. Dmitrenko, *Experimental observation of the tunnel effect for Cooper pairs with the emission of photons*, Sov. Phys. JETP **21**, 650 (1965).
3. D. N. Langenberg, D. J. Scalapino, B. N. Taylor, and R. E. Eck, *Investigation of microwave radiation emitted by Josephson junctions*, *Phys. Rev. Lett.* **15**, 294 (1965). Erratum *Phys. Rev. Lett.* **15**, 842 (1965).
4. A. J. Leggett, *Prospects in ultralow temperature physics*, *J. Phys. Colloq.* **39**, C6-1264-C61269 (1978).
5. A. O. Caldeira and A. J. Leggett, *Influence of dissipation on quantum tunneling in macroscopic systems*, *Phys. Rev. Lett.* **46**, 211 (1981).
6. A. O. Caldeira and A. J. Leggett, *Influence of damping on quantum interference: An exactly soluble model*, *Phys. Rev. A* **31**, 1059 (1985).
7. J. M. Martinis, M. H. Devoret, and J. Clarke, *Experimental tests for the quantum behavior of a macroscopic degree of freedom: The phase difference across a Josephson junction*, *Phys. Rev. B* **35**, 4682 (1987).
8. J. M. Martinis, M. H. Devoret, and J. Clarke, *Energy-level quantization in the zero-voltage state of a current-biased Josephson junction*, *Phys. Rev. Lett.* **55**, 1543 (1985).
9. I. M. Dmitrenko, G. M. Tsoi, and V. I. Shnyrkov, *Macroscopic quantum tunneling in a system with dissipation*, *Sov. J. Low Temp. Phys.* **8**, 330 (1982) [*Fiz. Nizk. Temp*. **8**, 660 (1982)].
10. I. M. Dmitrenko, V. A. Khlus, G. M. Tsoi, and V. I. Shnyrkov, *Study of quantum decays of metastable current states in RF SQUIDs*, *Sov. J. Low Temp. Phys.* **11**, 77 (1985) [*Fiz. Nizk. Temp*. **11**, 146 (1985)].
11. I. M. Dmitrenko, G. M. Tsoi, and V. I. Shnyrkov, *Quantum decay of metastable current states of a superconducting interferometer. Proceedings of 2nd meeting "Quantum metrology and fundamental physical constants"*, NPO VNIIM (1985), p. 81.
12. V. I. Shnyrkov, G. M. Tsoi, D. A. Konotop, and I. M. Dmitrenko, *Anomalous behaviour of RF-SQUIDs with S-c-S contacts of small area*, in: H. Koch and H. Lübbig, (eds.), *Single-Electron Tunneling and Mesoscopic Devices*, *Springer Series in Electronics and Photonics*, Springer, Berlin, Heidelberg (1992), Vol. 31, p. 211.
13. Y. Nakamura, Yu. A. Pashkin, and J. S. Tsai, *Coherent control of macroscopic quantum states in a single-Cooper-pair box*, *Nature* **398**, 786 (1999).
14. T. P. Orlando, J. E. Mooij, Lin Tian, Caspar H. van der Wal, L. S. Levitov, Seth Lloyd, and J. J. Mazo, *Superconducting persistent-current qubit*, *Phys. Rev. B* **60**, 15398 (1999).